# DTIP of MEMS & MOEMS

Stresa, Italy, 26-28 April 2006

# CONTACTLESS THERMAL CHARACTERIZATION METHOD OF PCB-S USING AN IR SENSOR ARRAY

Gy. Bognár[1], V. Székely[1], M. Rencz[1,2]

[1] Budapest University of Technology, Hungary, <bognar|szekely|rencz>@eet.bme.hu
[2] MicReD, Budapest, Hungary, rencz@micred.com

**ABSTRACT**

In this paper the feasibility study of an IR sensor card is presented. The methodology and the results of a quasi real-time thermal characterization tool and method for the temperature mapping of circuits and boards based on sensing the infrared radiation is introduced.

With the proposed method the IR radiation-distribution of boards from the close proximity of the sensor card is monitored in quasi real-time. The proposed method is enabling *in situ* IR measurement among operating cards of a system e.g. in a rack.

## 1  INTRODUCTION

The elevated temperature of different packaged electronic devices like digital processors, high power amplifier, high power switches, etc. demands careful temperature-aware design methodology and electro-thermal simulations of PCBs. The results of different electro-thermal simulations and *modeling* in most of the cases give the real results and consider the coupled effects of the real surroundings of these cards and other dissipation elements in an operating system. However the simulation time may take hours, and different systems, different surroundings should be simulated again and again.

In our expectation by using contactless temperature measurement procedure the heat distribution and the places of high dissipation elements on an operating PCB board (PCI or AGP cards in a rack-house of a PC) can be measured and localized.

There are several contactless temperature measurement methods, but the price and usability determine the application area.[1] The different types of thermographic cameras are not just too expensive but it is impossible to insert them between the cards of a system. So the heat distribution can not be visualized this way. An additional problem is that cooled type IR cameras need liquid gases during the operation. [2]

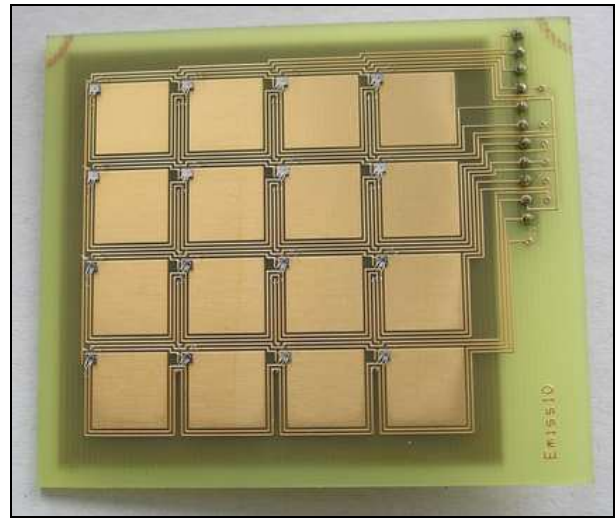

Figure 1. – The sensor card before black painting

The uncooled IR detectors are mostly based on pyroelectric materials or microbolometer technology. [3] Thanks to the MEMS technology small sensor array can be realized on a silicon die similar to CCD sensors. The only difference is in the sensing method. The CCD sensors are sensitive for the near infrared radiation as well, but they are used mainly for night vision application.

To determine the temperature of a distant object the sensing of the far infrared radiation is needed if the temperature changes of the devices are below 100°C. The main advantage of these sensors is the relatively small (in the 10ms range) time constant. The disadvantage is that additional optical elements (not glasses) are needed to get a relatively wide viewing angle. With these additional elements however the all IR sensor system may not be placed between two cards.

By using built-in temperature sensors integrated within the used integrated circuits only the temperature of the silicon die can be measured. The temperature of the packages and the PCB can not be determined, neither the heat distribution along the packages or the board.

In this paper a novel IR measurement system is introduced. In our solution the sensor card of the system will be placed between two cards, and the heat distribution of one of these PCB panels will be determined and visualized on a display. To check the feasibility of this methodology firstly a small resolution device has been developed for testing purposes.

## 2 THE MEASUREMENT SETUP

Between two cards inserted into two adjacent PCI slots there is about 10mm distance. Our device was realized on a thin (1.55 mm) PCB board with a special metallization pattern. In the first experiment a 4x4 matrix was created (Figure 1.). Each square shaped metal "pixel" on the card is aimed for absorbing IR radiation from the opposite region of the PCB. The temperature rise due to the absorbed IR radiation is sensed by using a thermal test-chip attached directly onto the copper plate of the "pixel". The thermal resistance between the central point of the pixel and the thermal test-chip had to be minimized therefore packaged devices could not be used for this purpose. For ultrasonic bonding the copper surface on the top side of the measurement card was covered by a 2um gold metallization layer. The thermal test chip provides an output frequency which is depending on the temperature rise of the "pixel". The design of the test chip used in this application is derived from the TMC family of TIMA and MicReD [4]. For minimizing the time constant of each pixel the thermal capacitance had to be decreased by selecting thinner board and thinner metallization layer (35um). The size of each pixel is 10mm × 10mm.

In this application the thermal sensor chips are addressable in the matrix. The frequency which is the signal proportional to temperature of the addressed pixel is available via a simple read-out electronics controlled by an ST7262 microcontroller. The communication between the PC and the electronics is realized by serial link (RS-232) communication.

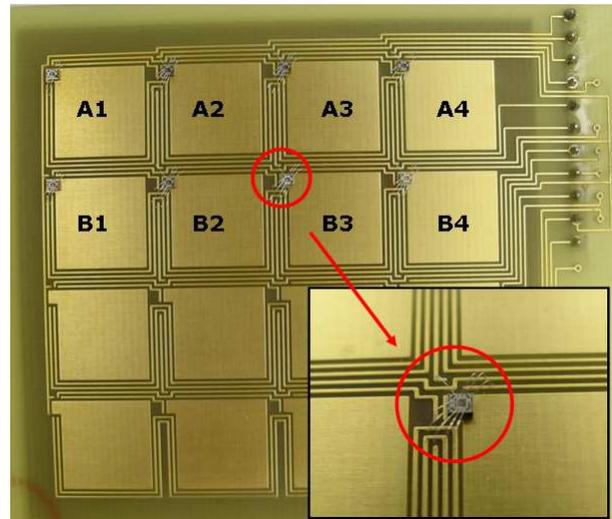

Figure 2. – The temperature sensor dies attached onto the pixels of the sensor card

The wiring of the measurement card is realized only on the top surface of the board without crossing. That's why some space (2…3mm) between the pixels had to be left. The width of the wires (180um) was selected by considering the bonding possibilities and the thickness of the bonding wire. The connector via the communication interface was placed on the right-hand side of the sensor card. For protecting the thermal dies from any damages the chips were covered by polymer protective layer. The all board was painted black to enable the higher absorption of heat radiation.

In the first approach 16 devices were placed and directly attached on the top surface of the measurement card by using an electrically conductive adhesive material which has very good thermal conductivity. However if the copper plate charges up electrically then the substrate voltage of the chip is increased. For this reason in certain conditions – electrical noise, supply voltage changes, etc. – the thermal die may not operate. To cope with this problem on the second board an electrically non- or wrong-conductive material was used to attach the dies (Figure 2.). However this type of adhesive material has significant thermal resistance. Consequently the time constant of each sensing pixel is increased. On the second boards 8 dies were placed.

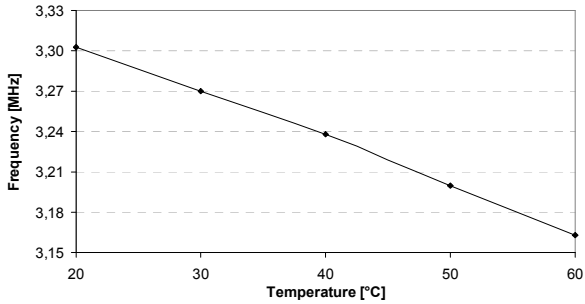

Figure 3. – The temperature vs. measured frequency diagram of the pixel A3

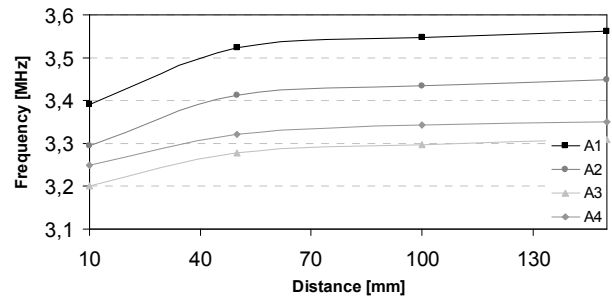

Figure 4. – The distance vs. frequency diagram

## 3 CALIBRATION

To enable calibration a large, pure copper black-painted plate was placed in front of the sensor card – on which 8 temperature sensing units were placed in a 2×4 matrix form. By using a thermostat unit the plate was heated up from the room-temperature to 60°C degree in 10°C steps. During the measurement the output frequency of the sensing unit was recorded which is related to the temperature of the pixel plate. The ambient temperature during all the measurements was 21 °C.

In every heating up step the sensor card was removed until the all black-painted plate reached the adjusted temperature. Then the sensor card was put in front of the black-body, exactly 10 mm far from the plate and in every minute the output frequencies as temperature dependent variables were recorded.

In Figure 3. the calibration result can be seen for the pixel A3. The approximately 100 kHz difference represents an almost 40°C temperature change.

In the second measurement the black-painted plate was heated up to 50°C and the measurement card was put from 10 mm to 150 mm far from the plate. The sensed temperature vs. distance diagram can be seen in figure 4. It can be seen that the measured temperature rise shows strong dependence on the distance between the heated and the sensor boards. It means that without calibration process only the place of the highest dissipation elements can be localized but the accurate temperature of the devices attached onto the measured board can not be determined without a preceding calibration process.

In figure 5. the sensed frequency – which represents the temperature rise vs. time function can be seen when the plate that is heated up to 60°C was placed at a 10 mm distance from the sensor card.. It can be recognized that the system reaches the near equilibrium state after 240 seconds. This unfortunately means that fast temperature changes can not be followed by using this method. Because of the physical dimension of the pixels – which means high thermal capacitances as well – we can not reach the 10ms time-constant of the integrated pyroelectric types sensors.

Otherwise if the time constant of the pixels can be decreased significantly (into the range of seconds) it will be appropriate for quasi real-time temperature distribution measurement.

Our initial experimental setup already proved the feasibility of this approach.

## 4 MEASUREMENT RESULT

During our third measurement a BD245C type transistor was used as a heat source. A real board was modeled by using a black painted, wrong heat-conductor paperboard. The transistor was placed in front of the A2 and B2 pixels, at a 10mm distance from the IR sensing board. A 700mW power step was applied to the transistor for 600 sec. The T3Ster [5] tester was used for driving the transistor and measuring the temperature change of it [7]. It can be seen that the A2 and B2 pixels sensed the biggest temperature changes. Table 1 shows the temperature change of the pixels. The measurement setup can be seen in Figure 6.

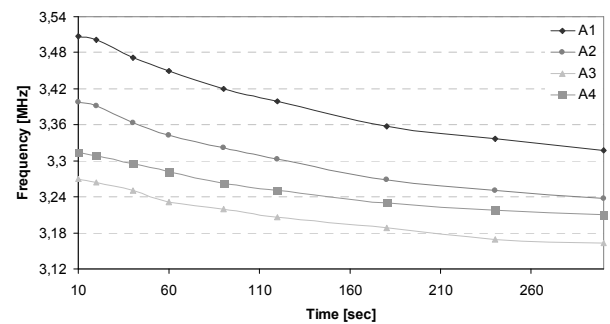

Figure 5. – The time vs. measured frequency function

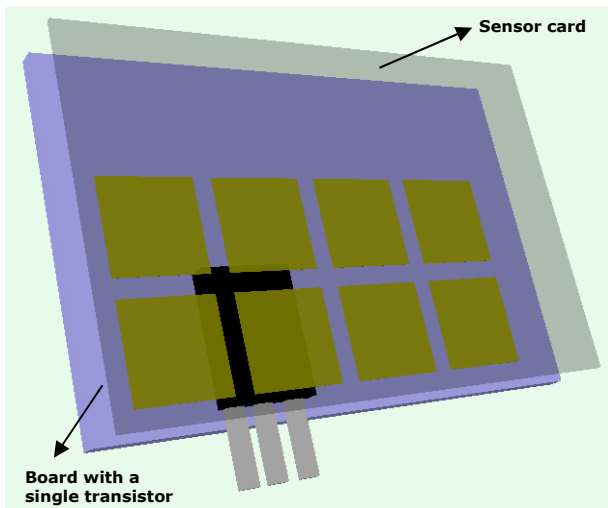

Figure 6. – The measurement setup

The measured temperature changes of the pixels determine exactly the location of the main dissipator element – in this case the used transistor. It proves the feasibility to determine the locations of the main dissipator elements.

Figure 7 shows that the dissipator element was in front of the B2 and A2 pixels during the measurement. The greatest difference between the temperature data of the pixels was 3.5 °C, and the B2 pixel reached the 40.4 °C. It can be seen that at the end of the measurement the all IR sensing card warmed up in some extent. Of course the pixels did not come to the same temperature. It is caused because of the finite thermal conductivity between the pixels and additionally the dissipator elements on the measured board radiate not only to the opposite pixels of the sensor card but the neighborhoods of it as well.

## 5 SIMULATION

A few number of simulations have been done for cross-verification purposes and to demonstrate the applicability of the sensor card. Steady-state solutions of our measuring system in a complete microATX type desktop rack system considering the heat radiation was done by using the FloTherm 5.1 simulation tools [8]. The built-in MicroATX template was completed with the black-painted sensor card placed between the *PCI 1* and *PCI 2* slots. The steady state results shows that between the cards inserted into PCI slots the air-flow is minimal. Onto one of the neighborhood card (PCI 1) a high dissipator devices was placed (10W) for representing the measurement setup described before. In the steady state solution this high dissipator devices reached almost 90°C without any active or passive cooling realizing a hotspot on the surface of a PCI card. The maximum temperature on the surface of the sensor card reached 46°C in the opposite place of the high dissipator element. This result shows correspondence of the results from the measurement, the difference between the simulated and the measured temperature elevations was caused by the heat radiation of the other neighborhood cards.

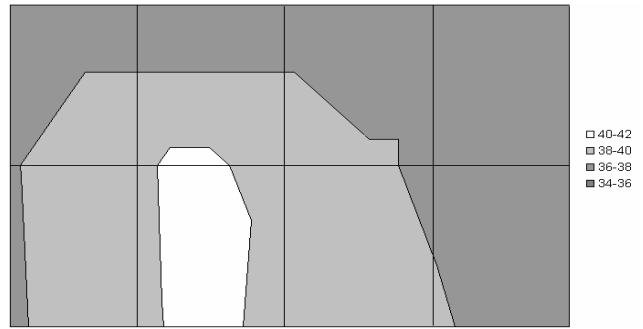

Figure 7. – The sensed heat distribution map

The temperature distribution on the sensor card inserted into two PCI cards and the place of the highest dissipator elements on the surface of the neighborhood card can be determined as it can be seen in Figure 8. The simulation showed the applicability of the novel IR sensor-card.

## 6 CONCLUSIONS

We have presented a contact-less method to localize thermal hot-spots. This method can be applicable for sensing the temperature distribution map of a card (PCI, AGP, …) in a dense rack system, where only a thin measuring board can be inserted between the cards during operation.

The disadvantage of our system is the high thermal resistance between the metal plate and the active area of the temperature sensor integrated circuit. For decreasing the time constant thinner PCB board, maybe back-drilled one under the metal plates and a better thermal conductor adhesive could be applied considering the ground bouncing and substrate noise effects as well.

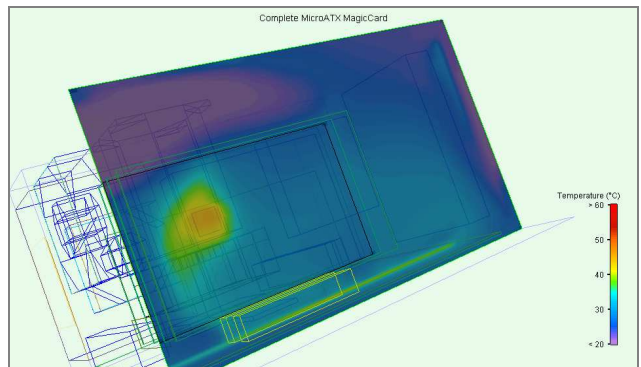

Figure 8. – The FloTherm simulation result

Application of only one MEMS IR sensor array integrated on one die isn't recommended to substitute the integrated circuits applied in our case, as only a small area of the measured PCB could be visualized. If we substitute our integrated circuits one at a time with several MEMS IR sensors (eg. stand-alone pyrometer or thermopile type sensor), which individually contain also a built in A/D converters (for digital output purposes), the time constant could be dramatically decreased.

**ACKNOWLEDGEMENT**

This work was supported by the INFOTHERM NKFP 2/018/2001 Project. The authors would like to thank the help of the Research Institute for Technical Physics and Materials Science of Hungary (MFA).